\documentclass[10pt,a4paper]{article}
\usepackage{geometry} 
\usepackage{amsmath}
\usepackage{amsthm}
\usepackage{amssymb}
\usepackage{amsfonts}
\usepackage[latin1]{inputenc}

\usepackage{cite}

\usepackage[usenames,dvipsnames]{color}

\def\be{\begin{equation}}
\def\ee{\end{equation}}
\def\bea{\begin{eqnarray}}
\def\eea{\end{eqnarray}}

\def\1{\'{\i}}                           
 
 \def\adsw{AdS$_\omega$\ }

\def\conm#1#2{\left[ #1,#2 \right]}

\def\k{\omega}

\def\R{T} 

\def\z{z}

\def\>#1{{\mathbf #1}}

\def\producto{\cdot}

\begin{document}


\begin{center}
\ 

\hfill

\bigskip

{\LARGE 
{\bf Curved momentum spaces from quantum  

\smallskip

(Anti-)de Sitter groups in (3+1) dimensions}}

\bigskip
\bigskip

{\sc A. Ballesteros$^1$, G. Gubitosi$^{2,3}$, I. Gutierrez-Sagredo$^1$, F.J. Herranz$^1$}

{$^1$ Departamento de F\1sica, Universidad de Burgos, 
E-09001 Burgos, Spain}

{$^2$ Radboud University, Institute for Mathematics, Astrophysics and Particle Physics, Heyendaalseweg 135, NL-6525 AJ Nijmegen, The Netherlands  
}

{$^{3}$ Dipartimento di Fisica, Universit\`a di Roma ``La Sapienza'', P.le A. Moro 2, 00185 Roma, Italy}
 
e-mail: {angelb@ubu.es,  g.gubitosi@science.ru.nl, igsagredo@ubu.es, fjherranz@ubu.es}

\end{center}


\begin{abstract}
Curved momentum spaces associated to the $\kappa$-deformation of the (3+1) de Sitter and Anti-de Sitter algebras are constructed as orbits of suitable actions of the dual Poisson-Lie group associated to the $\kappa$-deformation with non-vanishing cosmological constant. The $\kappa$-de Sitter and $\kappa$-Anti-de Sitter curved momentum spaces are separately analysed, and they turn out to be, respectively, half of the (6+1)-dimensional de Sitter space and half of a space with ${\rm SO}(4,4)$ invariance. Such spaces are made of  the momenta associated to spacetime translations and the `hyperbolic' momenta associated to boost transformations. The known $\kappa$-Poincar\'e curved momentum space is smoothly recovered as the vanishing cosmological constant limit from both of the constructions.
\end{abstract}

\section{Introduction}

It is now understood that in models of Planck-scale deformed special relativity (DSR), where the Planck energy plays the role of a second relativistic invariant besides the speed of light, momentum space  has a nontrivial geometry\cite{KowalskiGlikman:2003we, KowalskiGlikman:2002ft, Raetzel:2010je, AmelinoCamelia:2011bm, Mignemi:2016ilu}. In particular, its curvature can be non-vanishing and governed by the Planck energy \cite{Arzano:2014jua, Kowalski-Glikman:2013rxa}.

This realization recently allowed for a description of the phenomenological features of DSR as effects induced by the geometrical properties of momentum space,  dual to the effects  we are familiar  with in the context of general relativity. For example,  curvature of momentum space  induces an energy dependence in the distance covered by a free massless particle in a given time. This can be seen as the dual to the well-known redshift of energy in curved spacetime \cite{Amelino-Camelia:2013uya}. Another effect that can be ascribed to a nontrivial geometry of momentum space is the so-called dual-gravity lensing \cite{AmelinoCamelia:2011gy}, such that the direction from which a particle emitted by a given source reaches a faraway detector is energy dependent.
 
This kind of phenomenology raised considerable interest in the last two decades as it became clear that it could induce observable effects on  particles travelling over cosmological distances \cite{Ackermann:2009aa, AmelinoCamelia:1997gz, Amelino-Camelia:2016ohi, AmelinoCamelia:2009pg, Amelino-Camelia:2016fuh}.
However, the missing link to make the connection between properties of the momentum space and  cosmological observations more stringent was the definition of momentum space when spacetime is itself curved. DSR models are in fact modifications of special relativity, so that they need to be generalized to a curved-spacetime setting in order to be applicable to physical frameworks  where spacetime curvature cannot be neglected. Recently several proposals to extend DSR to curved spacetime scenarios have been put forward \cite{AmelinoCamelia:2012it, Barcaroli:2015xda, Barcaroli:2016yrl, Barcaroli:2017gvg, Amelino-Camelia:2014rga, Cianfrani:2014fia}, but it was generally believed that it would not be possible to look at the geometrical properties of the momentum space on its own, because the interplay between curvature of spacetime and of momentum space would make the phase space too much intertwined \cite{Marciano:2010gq}.

In recent  work we demonstrated  that in fact it is  possible to define momentum space in presence of a cosmological constant \cite{BGGHplb2017}.  The nontrivial interplay between curvature of spacetime and of momentum space is resolved by generalizing the momentum space so that  it contains the hyperbolic momenta associated to boosts besides the momenta associated to spacetime translations.
Our construction focussed on the mathematical framework of quantum deformed algebras, which  is known to be very effective in  studying  features of DSR models  in the flat spacetime case \cite{Gubitosi:2013rna, KowalskiGlikman:2004tz, AmelinoCamelia:2011nt}.  Specifically, we constructed the generalized momentum space associated to the $\kappa$-deformation of the de Sitter (dS) algebras in (1+1) and (2+1) dimensions. We obtained, respectively, (half of) a (2+1)-dimensional dS manifold and (half of) a (4+1)-dimensional dS manifold. The coordinates of these momentum spaces are the local group coordinates associated to spacetime translations and boosts and the two spaces are  the orbits of the corresponding dual Poisson-Lie group.
The construction crucially relied on the observation that spacetime translations and boosts play a similar role in the structure of the algebra and coalgebra of $\kappa$-dS in both (1+1) and (2+1) dimensions. Additionally, in (2+1) dimensions the rotation generator is not relevant to the construction of the generalized momentum space, since its role is limited to generating the isotropy subgroup of the  origin of the momentum space. 

While this previous work provided a crucial proof of concept, of course the physically relevant scenario is that with (3+1) spacetime dimensions. This is the case we focus on in this paper, where we construct the generalized momentum space of the (3+1)-dimensional $\kappa$-dS algebra. Moreover, we also present the  construction of the curved momentum space associated to the $\kappa$-deformation of the (3+1) Anti-de Sitter (AdS) algebra, which follows the same lines and --as it was conjectured in~\cite{BGGHplb2017}-- leads to a different manifold. Also the nontrivial features that arise in the (3+1) construction with respect to the (2+1) case for both quantum dS and AdS symmetries are emphasized.

We find it useful to begin in Section \ref{kP} with a summary of the construction of the curved momentum space associated to the $\kappa$-deformed Poincar\'e algebra in (3+1) dimensions. While this is a repetition of known results, it allows us to present   the framework that will be adopted in the following  within a context that should be more familiar to the reader. Concretely, we construct the dual Poisson-Lie group associated to the $\kappa$-Poincar\'e algebra and show that the orbits of this group define a curved momentum space manifold, which is half of a (3+1)-dimensional dS space.

Section \ref{kAds} is devoted to presenting the (3+1)-dimensional $\kappa$-(A)dS algebra and its dual Poisson-Lie group. The two cases with positive and negative cosmological constant can be treated in a unified framework, which also admits the $\kappa$-Poincar\'e case as the vanishing cosmological constant limit of either of $\kappa$-(A)dS.  We emphasize that the spacetime translation sector and the boosts enter in a very similar way in the structure of the  deformed coalgebra.  Moreover, the fact that a momentum space relying only on  the spacetime translations cannot be constructed is made apparent by the fact that the  coalgebra of translations does not close on its own, but it contains generators of the Lorentz sector. These two features are  analogous to what we had observed for the lower-dimensional models studied in our previous paper (the mixing between the translation sector and the Lorentz sector being somewhat less invasive in the (1+1) case, since it only concerns the algebra sector). A new feature that arises in the (3+1)-dimensional scenario  is the fact that the rotations sector is no longer undeformed at the coalgebraic level. 

In Sections \ref{AdSmomspace} and \ref{dSmomspace} we construct the generalized momentum spaces associated to, respectively, the $\kappa$-AdS and $\kappa$-dS algebra.  Again, for each algebra the momentum space is generated by the orbits of the dual Poisson-Lie group, and its coordinates are given by the local group coordinates associated to spacetime translations and boosts. However, while, as we mentioned, the two algebras and their dual Poisson-Lie groups admit a unified description, this is not the case for the two momentum spaces. This is because one needs to adapt the construction in order to deal with the imaginary quantities that arise when dealing with the two different signs of the cosmological constant.  As a consequence, the generalized momentum spaces that we find have different geometrical properties. The $\kappa$-AdS algebra has a momentum space that is half of a  ${\rm SO}(4,4)$ quadric. The  $\kappa$-dS algebra has a momentum space that is half of a (6+1)-dimensional dS manifold.
Similarly to what we had found in the lower-dimensional models, rotations do not enter in the construction of the momentum space, and only generate the isotropy subgroup of the origin.

The concluding Section \ref{conclusions} summarizes our findings and additional insights on the nontrivial properties of rotations that emerge in  (3+1) dimensions are included into an Appendix.


\section{The $\kappa$-Poincar\'e momentum space}\label{kP}

The aim of this Section is to summarize the construction of the curved momentum space corresponding to the (3+1) $\kappa$-Poincar\'e quantum algebra~\cite{Lukierskia, Lukierskib, Giller, Lukierskic}. We follow the presentation of~\cite{Kowalski-Glikman:2013rxa}, and  make use of the group-theoretical framework recently introduced in~\cite{BGGHplb2017}. We  show that the $\kappa$-Poincar\'e curved momentum space can be obtained as a specific orbit of the action of the dual $\kappa$-Poisson-Lie group on a (4+1)-dimensional ambient Minkowski space, and it turns out to be (half of) the (3+1) dS space. The rest of this paper is devoted to the generalization of this construction to the $\kappa$-deformed spacetime symmetries with non-vanishing cosmological constant.

The starting point for the construction is the Poisson version of the (3+1) $\kappa$-Poincar\'e algebra in the so-called bicrossproduct basis~\cite{Majid:1994cy}. This is given by the Poisson brackets  (the notation $\>P^2=P_1^2+P_2^2+P_3^2$ will be used hereafter for  vectors):
\be
\begin{array}{lll}
\left\{ J_a,J_b\right\}=\epsilon_{abc}J_c ,& \quad \left\{  J_a,P_b\right\}=\epsilon_{abc}P_c , &\quad
\left\{  J_a,K_b\right\}=\epsilon_{abc}K_c , \\[2pt]
\displaystyle{
  \left\{ K_a,P_0\right\}=P_a  } , &\quad\displaystyle{  \left\{ K_a,K_b\right\}=-\epsilon_{abc} J_c    } ,    &\quad\displaystyle{   \left\{ P_0,J_a\right\}=0  } , 
\\[2pt]
\left\{ P_0,P_a\right\}= 0 , &\quad   \left\{ P_a,P_b\right\}=0 , &\quad   \\[2pt]
\multicolumn{3}{l}
{\displaystyle      {   \left\{ K_a, P_b \right\} = \delta_{ab} \left( \frac{1}{2z} \left(1-e^{-2z P_0} \right) +\frac{z}{2} \>P^2 \right)- z P_a P_b  } \, ,} 
\end{array}
\label{poispoin}
\ee
together with the compatible (as a Poisson algebra homomorphism) coproduct map:
\begin{eqnarray}
&& \Delta_z ( P_0 ) = P_0 \otimes 1+1 \otimes P_0 ,\nonumber\\
&& \Delta_z ( P_a ) =  P_a \otimes 1+e^{-z P_0}  \otimes P_a  ,\label{copoinc} \\
&&  \Delta_z ( J_a ) =   J_a \otimes 1+1 \otimes J_a , \nonumber\\
&& \Delta_z ( K_a ) =  K_a \otimes  1+e^{-z P_0}  \otimes K_a+z\, \epsilon_{abc}   P_b \otimes J_c  \,.
\nonumber
\end{eqnarray}
We define $z=1/\kappa$ as the (Planck scale) quantum deformation parameter, with $a,b,c=1,2,3$  (sum over repeated indices is assumed) and $\epsilon_{abc}$ is the totally antisymmetric tensor, with $\epsilon_{123}=1$. As usual, the generators $P_a$, $J_a$ and $K_a$ denote respectively  translations, rotations and boosts.

This quantum deformation of the Poincar\'e algebra induces a deformed Casimir function  ${\cal C}_z$ for the Poisson algebra~\eqref{poispoin}, given by:
\be
 {\cal C}_z=\frac{4}{z^2}\sinh^2(zP_0/2) - e^{z P_0}\>P^2=\frac{2}{z^2}\bigl[  \cosh(zP_0)-1\bigr]- e^{z P_0}  \>P^2.
\label{casdeformeda}
\ee
This deformed Casimir constitutes the keystone for the interpretation of the $\kappa$-Poincar\'e algebra (in bicrossproduct basis) as the modified kinematical symmetry underlying  deformed dispersion relations arising in the quantum gravity context~\cite{KowalskiGlikman:2002ft, Kowalski-Glikman:2013rxa}. \footnote{A different choice of basis for the $\kappa$-Poincar\'e algebra would result in a different Casimir and thus characterise a different dispersion relation and associated kinematical symmetries.} It is also worth recalling that a second invariant $ {\cal W}_z$ does exist for the Poisson structure~\eqref{poispoin}, which is just a deformed analogue of the square of the modulus of the Pauli-Lubanski four-vector:
$$
 {\cal W}_z
=\left(  \cosh(zP_0)-  \frac {z^2}4 \, e^{z P_0}   \>P^2\right)W^2_{z,0}-\>{W}_z^2  \,  ,
$$
where the deformed components are:
$$
 W_{z,0}=e^{\frac z 2 P_0}\,  \>J\producto\>P ,
\qquad
 W_{z,a}=-  J_a \, \frac{\sinh(z P_0)}{z}+e^{z P_0}\epsilon_{abc} \left( K_b + \frac z 2 \, \epsilon_{bkl}J_k P_l \right)P_c   \, .
$$
We would like to stress that all the expressions included in this paper are analytic in the deformation parameter $z$, and the non-deformed limit $z\to 0$ always gives rise to the usual (3+1) relativistic symmetries. 

When dealing with Hopf algebra kinematical symmetries, the coproduct can be interpreted as the composition law for observables. In particular, the coproduct~\eqref{copoinc} is such that the $\kappa$-deformation induces a nonlinear composition rule for momenta in interaction vertices. As we are going to show, it is because of this  deformed composition rule that curvature in the $\kappa$-Poincar\'e momentum space emerges. In more technical terms, the curvature of the momentum space arises as a consequence of the non-cocommutativity of the coproduct map for the translation generators, and such a curved momentum space can be explicitly constructed as follows.

Firstly, the non-cocommutativity of the $\kappa$-translations can be characterized by writing the skew-symmetric part of the first-order deformation (in $z$) of the coproduct. Namely, if we write the coproduct~\eqref{copoinc} for the translation generators as a power series expansion in terms of the deformation parameter $z$ we obtain
\be
\Delta_z ( P_0 ) = P_0 \otimes 1+1 \otimes P_0,\qquad 
\Delta_z ( P_a ) =  P_a \otimes 1+1  \otimes P_a -z\, P_0  \otimes P_a + o[z^2],
\nonumber
\ee
and the skew-symmetrization of the first-order term in $z$ of the previuos expresions gives rise to the map
 \begin{align}
\delta(P_0)&=0, \nonumber\\
\delta(P_1)&=z \left(P_1 \wedge P_0 \right) , \nonumber\\
\delta(P_2)&=z \left(P_2 \wedge P_0  \right) ,\nonumber\\
\delta(P_3)&=z \left(  P_3 \wedge P_0 \right),\nonumber
\end{align}
which is called  the cocommutator map and it endows the Poincar\'e algebra $g$ with a Lie bialgebra structure $\delta: g\rightarrow g\otimes g$. Moreover, $\delta$ characterizes the Hopf algebra deformation through the first-order information it encodes  (see~\cite{CP,majid,dualJPA} and references therein for details).  

Secondly, the dual $\delta^\ast: g^\ast\otimes g^\ast\rightarrow g^\ast$ of the cocommutator map defines the Lie algebra $g^\ast$ of the so-called dual Poisson-Lie group $G^\ast$. In the $\kappa$-Poincar\'e case, if we denote by  $\{X^0,X^1,X^2,X^3\}$  the  generators in $g^\ast$ dual to, respectively, $\{P_0,P_1,P_2,P_3\}$, their dual Lie brackets are given by:
\be
\conm{X^0}{X^i}=-z\,X^i,
\qquad
\conm{X^i}{X^j}=0, 
\qquad i,j=1,2,3.
\label{kM}
\ee
Note that when the deformation parameter vanishes, $z\to 0$, then all the coproducts are cocommutative  ($\Delta_0(Y)=Y\otimes 1 + 1 \otimes Y$). As a consequence, $\delta$ vanishes and the dual Lie algebra (and group) is Abelian. It is also worth recalling that the dual Lie algebra of the translations sector given by~\eqref{kM} is just the so-called $\kappa$-Minkowski spacetime~\cite{kMinkowski,bicross2,kZakr,LukR}.

\subsection{Dual Poisson-Lie group and curved momentum space}

The dual Poisson-Lie group $G^\ast$ can be explicitly constructed starting from the 5-dimensional faithful representation $\rho$ of the dual Lie algebra $g^\ast$, given by:
\be
\begin{aligned}
&\rho(X^0)=z \left( 
\begin{array}{ccccc}
0 & 0 & 0 & 0 & 1\\
0 & 0 & 0 & 0 & 0\\
0 & 0 & 0 & 0 & 0\\
0 & 0 & 0 & 0 & 0\\
1 & 0 & 0 & 0 & 0
\end{array}
\right)\,,  \quad 
\rho(X^1)=z\left( 
\begin{array}{ccccc}
0 & 1 & 0 & 0 & 0\\
1 & 0 & 0 & 0 & 1\\
0 & 0 & 0 & 0 & 0\\
0 & 0 & 0 & 0 & 0\\
0 & -1 & 0 & 0 & 0
\end{array}
\right)\,,  
\\
&\rho(X^2)=z \left( 
\begin{array}{ccccc}
0 & 0 & 1 & 0 & 0\\
0 & 0 & 0 & 0 & 0\\
1 & 0 & 0 & 0 & 1\\
0 & 0 & 0 & 0 & 0\\
0 & 0 & -1 & 0 & 0
\end{array}
\right) \,, \quad 
\rho(X^3)=z\left( 
\begin{array}{ccccc}
0 & 0 & 0 &1 & 0\\
0 & 0 & 0 & 0 & 0\\
0 & 0 & 0 & 0 & 0\\
1 & 0 & 0 & 0 & 1\\
0 & 0 & 0 & -1 & 0
\end{array}
\right)\,. 
\label{prep2}\end{aligned}
\ee
If $\{p_0,p_1,p_2,p_3 \}$ are the local dual group coordinates associated, respectively,  to $\{X^0,X^1,X^2,X^3\}$,
then the dual Lie group element can be constructed through the exponentiation:
$$
G^\ast (p_0,p_1,p_2,p_3)=\exp \left(p_1 \rho(X^1) \right) \exp \left(p_2 \rho(X^2) \right) \exp \left(p_3 \rho(X^3) \right) \exp \left(p_0 \rho(X^0) \right),
$$
which explicitly reads:
\be
G^\ast  (p)=
\left( 
\begin{array}{ccccc}
\cosh(z p_0) \, +\frac{1}{2}\,e^{z\,p_0}\,z^2\, \bar{p}^2 & z p_1 &z p_2 & z p_3  & \sinh(z p_0) \, +\frac{1}{2}\,e^{z\,p_0}\,z^2\, \bar{p}^2\\
e^{z\,p_0}\,z p_1& 1 & 0 &0 & e^{z\,p_0}\,z p_1\\
e^{z\,p_0}\, z p_2 & 0 & 1 & 0 & e^{z\,p_0}\, z p_2\\
e^{z\,p_0}\,z p_3& 0 & 0 &1 & e^{z\,p_0}\,z p_3\\
\sinh(z p_0) \, -\frac{1}{2}\,e^{z\,p_0}\,z^2\, \bar{p}^2 & -z p_{1} & -z p_{2} & -z p_3 & \cosh(z p_0) \, -\frac{1}{2}\,e^{z\,p_0}\,z^2\, \bar{p}^2
\end{array}
\right),
\label{gpoincare}
\ee
where we have defined $\bar{p}^2=p_1^2 + p_2^2 + p_3^2$.

The significance of the dual Poisson-Lie group relies on the fact that the coproduct~\eqref{copoinc} is just the group law for $G^\ast$  (see~\cite{dualJPA} for details). In fact, if we multiply two matrices of the type~\eqref{gpoincare} we get another group element
$$
G^\ast  (p'')=G^\ast  (p) \cdot G^\ast  (p')\,.
$$
It can be straightforwardly checked that the group law $p''=f(p,p')$ reads:
\be
p_0''=p_0 + p_0', \qquad
p_i''=p_i+  e^{-{z} p_0} \, p'_i, 
\label{grouplaw}
\ee
which is consistent with~\eqref{kM} in the sense that $X^0$ generates a dilation and the $X^i$ generators correspond to (dual) translations.

Now, by making use of the Poisson version of the quantum duality principle~\cite{Dri,STS,GC}, the group multiplication law~\eqref{grouplaw} can be immediately rewritten in algebraic terms as a comultiplication map $\Delta_z$ through the identification of the two copies of the dual group coordinates as 
\be
p\equiv p \otimes 1,
\qquad
p'\equiv 1\otimes p.
\label{alg}
\ee
In this algebraic  language, the multiplication law for the group $G^\ast$ can be written as a coproduct  in the form:
\be
\Delta_z(p_0)\equiv p_0''=p_0 \otimes 1 + 1 \otimes p_0, \qquad
\Delta_z(p_i)\equiv p_i''=p_i\otimes 1 +  e^{-{z} p_0} \otimes p_i, 
\qquad i=1,2,3\,.
\label{cokappap}
\ee
This coproduct is just the one for the translation sector of the $\kappa$-Poincar\'e algebra in bicrossproduct basis once the following identification between the dual group coordinates and the generators of the $\kappa$-Poincar\'e algebra is performed:
$$
p_0\equiv P_0, \qquad
p_1\equiv P_1, \qquad
p_2\equiv P_2, \qquad
p_3\equiv P_3.
\label{id31p}
$$
Note that if a different ordering is chosen for the exponentials, one would recover the coproducts corresponding to a different basis of the $\kappa$-Poincar\'e algebra.
Moreover, the unique Poisson-Lie structure on $G^\ast$ that is compatible with the coproduct~\eqref{cokappap} and has the undeformed Poincar\'e Lie algebra as its linearization is given by the $\kappa$-Poincar\'e Poisson brackets for the translation sector.

Under this approach, the $\kappa$-Poincar\'e momentum space admits a straightforward geometric interpretation~\cite{Kowalski-Glikman:2013rxa}. The entries of the fifth column in $G^\ast$ can be rewritten as the following $S_i$ functions:
\be
\begin{aligned}
S_0&=\sinh(z p_0) \, +\frac{1}{2}\,e^{z p_0}\,z^2\, \bar{p}^2,  \\
S_1&= z\,   e^{z p_0} \, p_1 ,  \\
S_2&= z\,e^{z p_0}  \, p_2    , \\
S_3&= z\, e^{z p_0}   \,  p_3 ,  \\
S_4&= \cosh(z p_0) \, -\frac{1}{2}\,e^{z p_0}\,z^2\, \bar{p}^2\, .
\end{aligned}  
\label{ambkp}
\ee
Surprisingly enough, these satisfy the defining relation of the (3+1)-dimensional dS space:
$$
-S_0^2 + S_1^2 + S_2^2 + S_3^2 +S_4^2=1.
$$
This means that the $\kappa$-Poincar\'e momentum space parametrized by the ambient coordinates $(S_0,S_1,S_2,S_3,S_4)$ can be obtained as the orbit  arising from a linear action of the Lie group matrix $G^\ast(p)$ onto a (4+1)-dimensional ambient Minkowski space and passing through the point $(0,0,0,0,1)$. Namely:
\be
G^\ast\cdot
(0,0,0,0,1)^T=(S_0,S_1,S_2,S_3,S_4)^T\,.
\label{action}
\ee
Moreover, the ambient coordinates fulfil the condition:
$$
S_0+S_4=e^{z\,p_0}>0,
$$
which means that only half of the (3+1)-dimensional dS space is generated through the action~\eqref{action}. We will denote this manifold as $M_{dS_4}$.  Note that in the limit $z\to 0$ the dual Lie group $G^\ast$ generated by~\eqref{kM} is Abelian.


\section{The $\kappa$-(A)dS algebra and its dual Poisson-Lie group} \label{kAds}

The aim of this paper is the generalization of the previous construction to the case with non-vanishing cosmological constant, by following the approach presented in~\cite{BGGHplb2017} for the (1+1)- and (2+1)-dimensional  dS cases. In this way, a global picture of the interplay between the cosmological constant and the Planck-scale deformation parameter $z=1/\kappa$ can be presented. In our approach the cosmological constant $\Lambda=-\omega$ is included as an explicit deformation parameter and all the expressions are  analytic both in terms of $\Lambda$ and $z$. The construction of the (3+1) $\kappa$-(A)dS algebra was first exposed in~\cite{BHMNplb2017}. It turns out to be much more complicated than its $\kappa$-Poincar\'e limit, which is recovered when $\Lambda\to 0$. 

Before getting into the  $\kappa$-deformed case, let us briefly revisit the so-called AdS$_\omega$  algebra as a unified way to describe the AdS, dS and Poincar\'e symmetries (see~\cite{BHMNplb2017}). In the   kinematical basis the \adsw Lie algebra
$\{P_0,P_a, K_a, J_a\}$  is defined by the brackets:
\be
\begin{array}{lll}
[J_a,J_b]=\epsilon_{abc}J_c ,& \quad [J_a,P_b]=\epsilon_{abc}P_c , &\quad
[J_a,K_b]=\epsilon_{abc}K_c , \\[2pt]
\displaystyle{
  [K_a,P_0]=P_a  } , &\quad\displaystyle{[K_a,P_b]=\delta_{ab} P_0} ,    &\quad\displaystyle{[K_a,K_b]=-\epsilon_{abc} J_c} , 
\\[2pt][P_0,P_a]=\k  K_a , &\quad   [P_a,P_b]=-\k \epsilon_{abc}J_c , &\quad[P_0,J_a]=0  \,.
\end{array}
\label{aa}
\ee

As we mentioned, the parameter $\k$ is related to the cosmological constant via $\omega=-\Lambda$. Therefore,  the one-parametric AdS$_\omega$ algebra reduces to  the AdS  Lie algebra $\mathfrak{so}(3,2)$  when  $\k>0$,  to the     dS  Lie algebra $\mathfrak{so}(4,1)$  when   $\k<0$, and  to  the  Poincar\'e Lie algebra   $\mathfrak{iso}(3,1)$ when  $\k=0$.

The algebra~\eqref{aa} has two Casimir invariants~\cite{cass}. The first one is quadratic and comes from  the Killing--Cartan form:
$$
{\cal C}
= P_0^2-\>P^2  +\k \left(   \>J^2-\>K^2\right) .
\label{axa}
$$
The second one is a fourth-order invariant:
$$
{\cal W}
=W^2_0-\>{W}^2+\k \left(\>J\cdot\>K \right)^2,\label{axb}
$$
where  $W_0=\>J\cdot\>P$ and $W_a=-  J_a P_0+\epsilon_{abc} K_b P_c$ are the components of the (A)dS analogue of the
Pauli--Lubanski  4-vector. We recall that in the Poincar\'e case $\k=0$ the invariant $W^2_0-\>{W}^2$ provides the square of the spin/helicity operator, which in the rest frame is proportional to the square of the angular momentum operator. It is worth emphasising that the presence of a non-vanishing $\k$ implies that the quadratic invariant (the one linked to dispersion relations) has a new contribution coming from the Lorentz sector of the \adsw algebra.

By making use of this unified description, the  three    (3+1)  Lorentzian  symmetric homogeneous
spaces  with constant   sectional  curvature  $\k$  are defined as  the coset spaces ${\mathbf
{AdS}}^{3+1}_\k \equiv {\rm  SO}_{\k}(3,2)/{\rm  SO}(3,1) $, namely:
\begin{itemize}
\item When $\k>0$ ($\Lambda<0$) we have the AdS spacetime ${\mathbf
{AdS}}^{3+1} \equiv  {\rm SO}(3,2)/{\rm  SO}(3,1)$.

\item When $\k<0$ ($\Lambda>0$) we get the dS spacetime ${\mathbf
{dS}}^{3+1} \equiv   {\rm  SO}(4,1)/{\rm SO}(3,1)$.

\item The Minkowski spacetime ${\mathbf
M}^{3+1} \equiv  {\rm  ISO}(3,1)/{\rm  SO}(3,1)$ arises when $\k=\Lambda=0$.
\end{itemize}
Explicit ambient space coordinates for these three maximally symmetric Lorentzian spacetimes can be obtained by making use of a suitable realization of the Lie groups obtained by exponentiation of a faithful representation of the \adsw Lie algebra (see~\cite{tallin,BHMNsigma,ahep} for details in the (2+1)-dimensional case). 

In the following Subsections  we  summarize the main aspects of the $\kappa$-deformation of the \adsw Poisson algebra by making use of this unified description, and then we construct its associated dual Poisson-Lie group. The construction and analysis of the associated curved momentum spaces are  performed separately for the AdS and dS cases, since their geometric properties turn out to be different.


\subsection{The $\kappa$-deformation of the (3+1) \adsw algebra} 

The generalization of the (3+1) Poisson $\kappa$-Poincar\'e algebra to the non-vanishing $\k$ case has been recently presented explicitly in~\cite{BHMNplb2017}. In particular, the deformed coproduct for the \adsw algebra reads:
\be
\begin{aligned}
\Delta_z (J_3 ) &=  J_3 \otimes 1 +1 \otimes J_3 , \\
\Delta_z ( J_1  ) &=    J_1 \otimes e^{z\sqrt{\k} J_3} +1 \otimes J_1 , \\
 \Delta_z ( J_2 ) &= J_2 \otimes e^{z\sqrt{\k} J_3}+1 \otimes J_2  , 
\end{aligned}
\label{rot}
\ee
\be
\begin{aligned}
\Delta_z ( P_0 ) &=  P_0 \otimes 1+1 \otimes P_0   ,
 \\
\Delta_z ( P_1 )  &= P_1\otimes   \cosh (z\sqrt{\k} J_3) +e^{-z P_0}  \otimes P_1 
- \sqrt{\k} K_2 \otimes     { \sinh (z\sqrt{\k} J_3) }    \\
&\qquad - z\sqrt{\k}  P_3  \otimes J_1 + z {\k} K_3 \otimes J_2  + z^2 \k \left(\sqrt{\k} K_1-P_2 \right)  \otimes J_1 J_2 e^{-z\sqrt{\k}  J_3}    \\
& \qquad - \frac 12 z^2 {\k}   \left( \sqrt{\k} K_2+P_1   \right)  \otimes      \left( J_1^2-J_2^2 \right)    e^{-z\sqrt{\k} J_3}   ,    \\
\Delta_z ( P_2 ) &= P_2\otimes  \cosh( z\sqrt{\k} J_3) +e^{-z P_0}  \otimes P_2
+ \sqrt{\k} K_1 \otimes    { \sinh (z\sqrt{\k} J_3) }   \\
& \qquad - z\sqrt{\k}  P_3  \otimes J_2 - z {\k} K_3 \otimes J_1  - z^2 \k \left(\sqrt{\k} K_2+P_1 \right)  \otimes J_1 J_2 e^{-z\sqrt{\k}  J_3}     \\
&\qquad  - \frac 12 z^2 {\k}    \left(    \sqrt{\k} K_1- P_2 \right)  \otimes     \left( J_1^2-J_2^2 \right)       e^{-z\sqrt{\k} J_3}  ,   \\
\Delta_z ( P_3 )  &= P_3  \otimes 1 +e^{-z P_0}  \otimes P_3  + z  \left( {\k}  K_2+\sqrt{\k}  P_1 \right)  \otimes J_1 e^{-z \sqrt{\k} J_3}    \\
&\qquad  -z    \left( {\k}  K_1-\sqrt{\k}  P_2 \right) \otimes  J_2 e^{-z \sqrt{\k}  J_3}  ,   
\end{aligned}
\label{transl}
\ee
\be
\begin{aligned}
\Delta_z ( K_1 ) &=   K_1\otimes     \cosh( z\sqrt{\k} J_3) +e^{-z P_0}  \otimes K_1 +  P_2 \otimes       \frac{ \sinh (z\sqrt{\k} J_3) } {\sqrt{\k}}    \\
& \qquad - z P_3  \otimes J_2 - z\sqrt{\k} K_3 \otimes J_1  - z^2  \left(\k K_2+\sqrt{\k}P_1 \right)  \otimes J_1 J_2 e^{-z\sqrt{\k}  J_3}     \\
&\qquad  - \frac 12 z^2   \left(  {\k}  K_1-  \sqrt{\k}  P_2 \right) \otimes  \left( J_1^2-J_2^2 \right)    e^{-z\sqrt{\k} J_3} ,    \\
\Delta_z ( K_2 )&=  K_2\otimes    \cosh (z\sqrt{\k} J_3) +e^{-z P_0}  \otimes K_2 -  P_1 \otimes      \frac{ \sinh (z\sqrt{\k} J_3) } {\sqrt{\k}}   \\
&\qquad + z P_3  \otimes J_1 - z\sqrt{\k} K_3 \otimes J_2  - z^2  \left(\k K_1-\sqrt{\k}P_2 \right)  \otimes J_1 J_2 e^{-z\sqrt{\k}  J_3}    \\
&\qquad +\frac 12 z^2 \left(\k K_2+\sqrt{\k}  P_1  \right)  \otimes   \left( J_1^2-J_2^2 \right) e^{-z \sqrt{\k}J_3}   ,  \\
\Delta_z ( K_3 )&=   K_3 \otimes 1+e^{-z P_0}  \otimes K_3 +  z    (\sqrt{\k}  K_1- P_2) \otimes J_1 e^{-z \sqrt{\k}  J_3}   \\
& \qquad + z  (\sqrt{\k}  K_2+P_1) \otimes  J_2 e^{-z \sqrt{\k} J_3}   .
\end{aligned}
\label{boosts}
\ee
 Notice that this coproduct is written in a `bicrossproduct-type' basis that generalizes the one corresponding to the (2+1) $\kappa$-AdS$_\omega$ algebra~\cite{BHMNsigma,ahep}. 

As it can be easily checked, the $\kappa$-Poincar\'e coproduct~\eqref{copoinc} is obtained from the above expressions in the limit $\k\to 0$. A direct comparison between both sets of expressions makes it evident that the degree of complexity of the $\kappa$-deformation is greatly increased when the cosmological constant $\omega$  is turned on. In fact, the $\kappa$-\adsw algebra can be thought of as a two-parametric deformation, which is ruled by a `quantum' deformation parameter $z=1/\kappa$ (the Planck scale) and a `classical' deformation parameter $\k=-\Lambda$ (the cosmological constant) which has a well-defined geometrical meaning. 
As we show in the following, the roles of the two deformation parameters are interchanged when the dual Poisson-Lie group is considered, in the spirit of the `semidualization' approach to (2+1) quantum gravity~\cite{MSsemid, OSsemid}.

There are several differences between the coproducts~\eqref{rot}-\eqref{boosts} and~\eqref{copoinc} that have to be emphasized. First, $\Delta_z(K)$ and $\Delta_z(P)$ are structurally similar when $\k\neq 0$, in contrast with~\eqref{copoinc}. Second,  translations in~\eqref{transl} do not close a Hopf subalgebra, since when $\k\neq 0$ the coproducts $\Delta_z(P)$ contain boosts and rotations as well. Finally, in the non-vanishing cosmological constant case the rotation sector~\eqref{rot}  is deformed, whilst in~\eqref{copoinc} all of the coproducts for $J_a$ are  primitive. In particular, the rotation generators have no longer all the same role in the coalgebra, signaling a departure from standard isotropy. Further comments on this are found below, as well as in the concluding section.\footnote{Some technical comments on the deformation~\eqref{rot} of the rotation subalgebra in the (3+1)-dimensional $\kappa$-\adsw algebra are presented in the Appendix. In fact, this last feature is quite different from other kinds of deformations that have been studied in the context of Planck-scale deformed symmetries 
and --to the best of our knowledge-- has not been appropriately emphasized in the literature.}
These three  features  are induced by the interplay between the cosmological constant and the quantum deformation, and the  first two will be essential for the construction of the curved momentum space when $\k\neq 0$.

We do not reproduce here the deformed brackets for the Poisson version of the $\kappa$-\adsw algebra which can be explicitly found in~\cite{BHMNplb2017}, since they are quite involved and strongly present non-linear expressions when compared to the \adsw Lie algebra~\eqref{aa}. It is worth stressing that, in contradistinction to~\eqref{poispoin}, the full Lorentz sector has now deformed Poisson brackets like, for instance 
\bea
&& \left\{ J_1,J_2 \right\}= \frac{e^{2z\sqrt{\k}J_3}-1}{2z\sqrt{\k} } - \frac{z\sqrt{\k}}{2} \left(J_1^2+J_2^2\right)  ,\qquad
\left\{ J_1,K_2 \right\}=  K_3 -z  \sqrt{\k} J_1 K_1   ,\cr
&&  \left\{ K_2, K_3 \right\}=-\frac12 J_1 \left( 1+e^{-2z \sqrt{\k}  J_3}\left[ 1+  {z^2\k }   \left( J_1^2 +J_2^2 \right) \right] \right) -z  \sqrt{\k}  K_1 K_3  ,
\nonumber
\eea
which show evident differences with respect to its $\k\to 0$ limit~\eqref{poispoin}, and where the role of the $J_3$ generator is neatly distinguished with respect to the one played by $J_1$ and $J_2$. 

The deformed quadratic Casimir for the $\kappa$-\adsw algebra reads:
\bea
{\cal C}_z\!\!\!&=&\!\!\! \frac 2{z^2}\left[ \cosh (zP_0)\cosh(z\sqrt{\k} J_3)-1 \right]+\k \cosh(z P_0) (J_1^2+J_2^2) e^{-z \sqrt{\k}  J_3}\nonumber\\
&& -e^{zP_0} \left( \mathbf{P}^2 +\k \mathbf{K}^2 \right)   \left[ \cosh(z\sqrt{\k}  J_3)+ \frac { z^2\k}2  (J_1^2+J_2^2)  e^{-z \sqrt{\k}  J_3} \right]\label{casa}\\
&&+2\k e^{zP_0}  \left[ \frac{\sinh(z \sqrt{\k}J_3)}{\sqrt{\k}}\R_3+ z \left( J_1\R_1 +J_2 \R_2+  \frac {z \sqrt{\k}}2 (J_1^2+J_2^2) \R_3 \right)  e^{-z \sqrt{\k}  J_3} \right],
\nonumber
\eea
where $\R_a=\epsilon_{abc} K_b P_c$.  Again, the $\k\to 0$ limit is just~\eqref{casdeformeda}, and comparing~\eqref{casa}  to its `flat' limit~\eqref{casdeformeda} gives a clear idea of the kind of deformation we are dealing with. This deformed invariant~\eqref{casa} is relevant in what follows, since it is connected to the deformed dispersion relation that can be deduced from the curved momentum space with cosmological constant that we are going to construct.


\subsection{The dual Poisson-Lie group}

Mimicking the procedure used in the previous Section, the first step for the construction of the curved momentum spaces of the $\kappa$-(A)dS algebras  is to obtain the cocommutator map $\delta$ associated to the $\kappa$-deformed coproduct map with non-vanishing cosmological constant. This can be found by extracting the first-order deformation in $z$ of the coproduct~\eqref{rot}-\eqref{boosts}, which has a skew-symmetric part given by~\cite{BHMNplb2017}:
 \begin{align}
\delta(P_0)&=0,\qquad \delta(J_3)=0 , \nonumber \\
\delta(J_1)&=z \sqrt{\k}   J_1 \wedge J_3  ,\qquad 
\delta(J_2)=z \sqrt{\k}   J_2 \wedge J_3   ,\nonumber \\
\delta(P_1)&=z \left(P_1 \wedge P_0 -\k   J_2  \wedge  K_3+  \k J_3 \wedge K_2 + \sqrt{\k}J_1 \wedge P_3 \right) , \nonumber\\
\delta(P_2)&=z \left(P_2 \wedge P_0 - \k  J_3  \wedge K_1+ \k J_1 \wedge K_3 + \sqrt{\k}  J_2 \wedge P_3 \right) ,\nonumber\\
\delta(P_3)&=z \left(  P_3 \wedge P_0- \k J_1\wedge  K_2  + \k J_2 \wedge K_1  - \sqrt{\k} J_1  \wedge P_1 - \sqrt{\k}  J_2  \wedge P_2 \right), \label{ac}\\
\delta(K_1)&=z \left(K_1 \wedge P_0 + J_2 \wedge P_3 - J_3 \wedge  P_2  +\sqrt{\k}  J_1 \wedge K_3  \right), \nonumber\\
\delta(K_2)&=z \left(   K_2 \wedge P_0+ J_3 \wedge P_1 - J_1  \wedge P_3 +\sqrt{\k} J_2 \wedge K_3   \right) ,\nonumber \\
\delta(K_3)&=z \left(  K_3 \wedge P_0 + J_1 \wedge P_2 - J_2  \wedge P_1  - \sqrt{\k} J_1   \wedge K_1  -\sqrt{\k}  J_2  \wedge  K_2 \right)  .
\nonumber
\end{align}
The differences between the (A)dS and Poicar\'e deformations which were mentioned in the previous Subsection leave their traces in the $\delta$ map. In particular, we stress two main features of the cocommutator~\eqref{ac}: firstly,  that $\delta(P)$ does not close a sub-Lie bialgebra since it includes the full Lorentz sector in the definition of the cocommutator; secondly, that $\delta(P)$ and $\delta(K)$ are structurally similar when $\k\neq 0$. The latter statement is just the footprint of a general  property of classical dS and AdS symmetries, that disappears in the Poincaré limit, since
the \adsw Lie algebra~\eqref{aa} with $\k\ne 0$ can be straightforwardly endowed with the following automorphism that interchanges the $P_a$ and $K_a$ generators:
\be
\tilde P_0=P_0,\qquad \tilde P_a= \sqrt {\k}\, K_a,\qquad \tilde K_a=-\frac{1}{\sqrt{\k}}\, P_a,\qquad \tilde J_a= J_a .
\label{auto}
\ee
 It can directly be checked that the transformed generators close the commutation rules (\ref{aa}), and that the cocommutator (\ref{ac}), coproduct (\ref{rot})-(\ref{boosts}) and deformed Poisson brackets given in~\cite{BHMNplb2017}   remain  in the same form, provided that  the deformation parameter $z$ is unchanged.
Therefore, when the cosmological constant does not vanish, translations and boosts have similar  algebraic properties and play complementary geometric roles (see~\cite{ahep} for a more detailed discussion in the context of homogeneous   spaces of worldlines).

As a consequence, the main idea introduced in~\cite{BGGHplb2017} for the construction of curved momentum spaces for the $\kappa$-(A)dS algebras in (2+1) dimensions becomes  fully applicable: when $\k\neq 0$ the momentum space has to be enlarged by including the angular momenta associated to the rotation symmetries and the `hyperbolic' momenta associated to boosts. This means that the momentum space  arises as the orbit of an appropriate action of the dual Poisson-Lie group $G^\ast_\k$, whose Lie algebra $g^\ast_\k$ is obtained by dualizing $\delta$. Namely:
\be
\begin{array}{lll}
\left[ R^1,R^2 \right]=0 ,& \quad \left[ R^1,R^3 \right]= z\sqrt{\k}  R^1   , &\quad
 \left[ R^2,R^3 \right]=z\sqrt{\k}  R^2, \\[2pt]
\left[ R^1, X^1 \right]= - z \sqrt{\k}  X^3 ,& \quad \left[R^1,  X^2 \right]= z L^3   , &\quad
\left[R^1,  X^3 \right]= -z (L^2-\sqrt{\k}   X^1)   , \\[2pt]
\left[ R^2, X^1 \right]= -z L^3  ,& \quad \left[ R^2 , X^2\right]=- z\sqrt{\k}   X^3   , &\quad
 \left[ R^2, X^3 \right]=z(L^1 + \sqrt{\k}   X^2)  , \\[2pt]
 \left[ R^3,X^1 \right]= z L^2  ,& \quad \left[ R^3,X^2 \right]= -z L^1     , &\quad
 \left[ R^3, X^3 \right]=0 , \\[2pt]
\left[ R^1 ,L^1 \right]= -z \sqrt{\k}  L^3  ,& \quad  \left[ R^1,L^2 \right]=- z {\k} X^3     , &\quad
 \left[ R^1,L^3 \right]=z (\sqrt{\k} L^1+ {\k} X^2)   , \\[2pt]
\left[ R^2 ,L^1 \right]=  z  {\k}  X^3  ,& \quad  \left[ R^2,L^2 \right]=- z \sqrt{\k} L^3     , &\quad
 \left[ R^2,L^3 \right]=z (\sqrt{\k} L^2- {\k} X^1)   , \\[2pt]
\left[ R^3 ,L^1 \right]=  -z  {\k} X^2  ,& \quad  \left[ R^3,L^2 \right]= z {\k} X^1     , &\quad
 \left[ R^3,L^3 \right]=0 , \\[2pt]
\left[ L^a ,X^0 \right]=  z L^a  ,& \quad  \left[ L^a,L^b \right]= 0     , &\quad
 \left[ L^a,X^b \right]=0 , \\[2pt]
\left[ X^a ,X^0 \right]=  z X^a  ,& \quad  \left[ X^a,X^b \right]= 0     , &\quad
 \left[ X^0,R^a \right]=0 \,,
 \end{array}
\label{ba}
\ee
where $\{X^0,X^1,X^2,X^3,L^1,L^2,L^3,R^1,R^2,R^3\}$ are dual to $\{P_0,P_1,P_2,P_3,K_1,K_2,K_3,J_1,J_2,J_3\}$, respectively.
Notice that (\ref{auto}) induces, through duality and provided that $\k\ne0$, the following automorphism for the generators of the dual Lie algebra $g^\ast_\omega$
 $$
 \tilde X^0=X^0,\qquad \tilde X^a=\frac 1{ \sqrt {\k}}\, L^a,\qquad \tilde L^a=- {\sqrt{\k}}\, X^a,\qquad \tilde R^a= R^a ,
 $$
which leaves the commutation relations (\ref{ba}) invariant  and  shows that the  $X^a$ and $L^a$ generators can be also interchanged at the dual Lie algebra level.

Next from  the expressions (\ref{ba})  we deduce that in $g^\ast_\k$ there exists a seven-dimensional solvable Lie subalgebra generated by:
\be
 [X^0, X^i]=-z  X^i    ,\quad
 [X^0, L^i]=-z  L^i  , \quad
 [X^i, L^j]=0  , \quad
[X^i, X^j]=0  , \quad
 [L^i, L^j]=0  , \quad
i=1,2,3. \label{genkM}
\ee
This subalgebra is $\k$-independent, and the dual of the rotation sector generates a three-dimensional solvable subalgebra:
$$
\left[ R^1,R^2 \right]=0 , \quad \left[R^1,R^3 \right]= z\sqrt{\k}  R^1   , \quad
 \left[ R^2,R^3 \right]=z\sqrt{\k} R^2\,.
 \label{dualrot}
$$
In the limit $\k\to 0$ this turns out to be Abelian, which is the dual counterpart of the fact that $\Delta_z(J_a)=\Delta_0(J_a)$ when the quantum deformation disappears. 

We stress that the first-order noncommutative $\kappa$-\adsw spacetime would be given by the dual of the translations sector, namely:
$$
 [X^0, X^i]=-z   X^i    ,\qquad
[X^i, X^j]=0  ,
\qquad i=1,2,3.
$$
This is indeed $\k$-independent but, as it was shown in~\cite{RossanoPLB,BHMNsigma}, when the all-orders quantum group is computed, the quantum spacetime with non-vanishing $\k$ is a nonlinear algebra whose higher-order contributions explicitly depend on the cosmological constant.


\section{The $\kappa$-AdS curved momentum space}\label{AdSmomspace}

In this Section and the following we separately analyze the $\kappa$-AdS and $\kappa$-dS dual Poisson-Lie groups and construct the associated momentum spaces, since their geometric properties are different.
The matrix representation for the Lie algebra~\eqref{ba} when $\k>0$ can be found to be
{\small
\be
D(X^0)=
z \left(
\begin{array}{cccccccc}
 0 & 0 & 0 & 0 & 0 & 0 & 0 & 1 \\
 0 & 0 & 0 & 0 & 0 & 0 & 0 & 0 \\
 0 & 0 & 0 & 0 & 0 & 0 & 0 & 0 \\
 0 & 0 & 0 & 0 & 0 & 0 & 0 & 0 \\
 0 & 0 & 0 & 0 & 0 & 0 & 0 & 0 \\
 0 & 0 & 0 & 0 & 0 & 0 & 0 & 0 \\
 0 & 0 & 0 & 0 & 0 & 0 & 0 & 0 \\
 1 & 0 & 0 & 0 & 0 & 0 & 0 & 0 \\
\end{array}
\right),
\qquad
D(X^1)=
z \left(
\begin{array}{cccccccc}
 0 & 1 & 0 & 0 & 0 & 0 & 0 & 0 \\
 1 & 0 & 0 & 0 & 0 & 0 & 0 & 1 \\
 0 & 0 & 0 & 0 & 0 & 0 & 0 & 0 \\
 0 & 0 & 0 & 0 & 0 & 0 & 0 & 0 \\
 0 & 0 & 0 & 0 & 0 & 0 & 0 & 0 \\
 0 & 0 & 0 & 0 & 0 & 0 & 0 & 0 \\
 0 & 0 & 0 & 0 & 0 & 0 & 0 & 0 \\
 0 & -1 & 0 & 0 & 0 & 0 & 0 & 0 \\
\end{array}
\right),
\nonumber
\ee
\be
D(X^2)=
z \left(
\begin{array}{cccccccc}
 0 & 0 & 1 & 0 & 0 & 0 & 0 & 0 \\
 0 & 0 & 0 & 0 & 0 & 0 & 0 & 0 \\
 1 & 0 & 0 & 0 & 0 & 0 & 0 & 1 \\
 0 & 0 & 0 & 0 & 0 & 0 & 0 & 0 \\
 0 & 0 & 0 & 0 & 0 & 0 & 0 & 0 \\
 0 & 0 & 0 & 0 & 0 & 0 & 0 & 0 \\
 0 & 0 & 0 & 0 & 0 & 0 & 0 & 0 \\
 0 & 0 & -1 & 0 & 0 & 0 & 0 & 0 \\
\end{array}
\right),
\qquad
D(X^3)=
z \left(
\begin{array}{cccccccc}
 0 & 0 & 0 & 1 & 0 & 0 & 0 & 0 \\
 0 & 0 & 0 & 0 & 0 & 0 & 0 & 0 \\
 0 & 0 & 0 & 0 & 0 & 0 & 0 & 0 \\
 1 & 0 & 0 & 0 & 0 & 0 & 0 & 1 \\
 0 & 0 & 0 & 0 & 0 & 0 & 0 & 0 \\
 0 & 0 & 0 & 0 & 0 & 0 & 0 & 0 \\
 0 & 0 & 0 & 0 & 0 & 0 & 0 & 0 \\
 0 & 0 & 0 & -1 & 0 & 0 & 0 & 0 \\
\end{array}
\right),
\nonumber
\ee
\be
D(L^1)=
z\,\sqrt{\omega }  \left(
\begin{array}{cccccccc}
 0 & 0 & 0 & 0 & -1 & 0 & 0 & 0 \\
 0 & 0 & 0 & 0 & 0 & 0 & 0 & 0 \\
 0 & 0 & 0 & 0 & 0 & 0 & 0 & 0 \\
 0 & 0 & 0 & 0 & 0 & 0 & 0 & 0 \\
 1 & 0 & 0 & 0 & 0 & 0 & 0 & 1 \\
 0 & 0 & 0 & 0 & 0 & 0 & 0 & 0 \\
 0 & 0 & 0 & 0 & 0 & 0 & 0 & 0 \\
 0 & 0 & 0 & 0 & 1 & 0 & 0 & 0 \\
\end{array}
\right),
\qquad
D(L^2)=
z\,\sqrt{\omega }  \left(
\begin{array}{cccccccc}
 0 & 0 & 0 & 0 & 0 & 1 & 0 & 0 \\
 0 & 0 & 0 & 0 & 0 & 0 & 0 & 0 \\
 0 & 0 & 0 & 0 & 0 & 0 & 0 & 0 \\
 0 & 0 & 0 & 0 & 0 & 0 & 0 & 0 \\
 0 & 0 & 0 & 0 & 0 & 0 & 0 & 0 \\
 -1 & 0 & 0 & 0 & 0 & 0 & 0 & -1 \\
 0 & 0 & 0 & 0 & 0 & 0 & 0 & 0 \\
 0 & 0 & 0 & 0 & 0 & -1 & 0 & 0 \\
\end{array}
\right),
\nonumber
\ee
\be
D(L^3)=
z\,\sqrt{\omega }  \left(
\begin{array}{cccccccc}
 0 & 0 & 0 & 0 & 0 & 0 & 1 & 0 \\
 0 & 0 & 0 & 0 & 0 & 0 & 0 & 0 \\
 0 & 0 & 0 & 0 & 0 & 0 & 0 & 0 \\
 0 & 0 & 0 & 0 & 0 & 0 & 0 & 0 \\
 0 & 0 & 0 & 0 & 0 & 0 & 0 & 0 \\
 0 & 0 & 0 & 0 & 0 & 0 & 0 & 0 \\
 -1 & 0 & 0 & 0 & 0 & 0 & 0 & -1 \\
 0 & 0 & 0 & 0 & 0 & 0 & -1& 0 \\
\end{array}
\right),
\qquad
D(R^1)=
z\,\sqrt{\omega }  \left(
\begin{array}{cccccccc}
 0 & 0 & 0 & 0 & 0 & 0 & 0 & 0 \\
 0 & 0 & 0 & 0 & 0 & 0 & 1 & 0 \\
 0 & 0 & 0 & 1 & 0 & 0 & 0 & 0 \\
 0 & 0 & -1 & 0 & 1 & 0 & 0 & 0 \\
 0 & 0 & 0 & 1 & 0 & 0 & 0 & 0 \\
 0 & 0 & 0 & 0 & 0 & 0 & 1 & 0 \\
 0 & 1 & 0 & 0 & 0 & -1 & 0 & 0 \\
 0 & 0 & 0 & 0 & 0 & 0 & 0 & 0 \\
\end{array}
\right),
\nonumber
\ee
\be
D(R^2)=
z\,\sqrt{\omega }  \left(
\begin{array}{cccccccc}
 0 & 0 & 0 & 0 & 0 & 0 & 0 & 0 \\
 0 & 0 & 0 & 1 & 0 & 0 & 0 & 0 \\
 0 & 0 & 0 & 0 & 0 & 0 & -1 & 0 \\
 0 & -1 & 0 & 0 & 0 & 1 & 0 & 0 \\
 0 & 0 & 0 & 0 & 0 & 0 & -1 & 0 \\
 0 & 0 & 0 & 1 & 0 & 0 & 0 & 0 \\
 0 & 0 & -1 & 0 & 1 & 0 & 0 & 0 \\
 0 & 0 & 0 & 0 & 0 & 0 & 0 & 0 \\
\end{array}
\right),
\qquad
D(R^3)=
z\,\sqrt{\omega }  \left(
\begin{array}{cccccccc}
 0 & 0 & 0 & 0 & 0 & 0 & 0 & 0 \\
 0 & 0 & 0 & 0 & 0 & -1 & 0 & 0 \\
 0 & 0 & 0 & 0 & -1 & 0 & 0 & 0 \\
 0 & 0 & 0 & 0 & 0 & 0 & 0 & 0 \\
 0 & 0 & -1 & 0 & 0 & 0 & 0 & 0 \\
 0 & -1 & 0 & 0 & 0 & 0 & 0 & 0 \\
 0 & 0 & 0 & 0 & 0 & 0 & 0 & 0 \\
 0 & 0 & 0 & 0 & 0 & 0 & 0 & 0 \\
\end{array}
\right).
\nonumber
\ee}
This faithful representation has been found by imposing that its $\omega\to 0$ limit should lead to a reducible representation in which the matrices~\eqref{prep2} can be obtained by suppressing appropriate rows and columns. After that, the rest of entries were found through direct computation by imposing the commutation rules~\eqref{ba} to hold.

If we denote as $\{p_0,p_1,p_2,p_3,\chi_1,\chi_2,\chi_3,\theta_1,\theta_2,\theta_3\}$ the local dual group coordinates that correspond, respectively, to $\{X^0,X^1, X^2, X^3, L^1,L^2, L^3, R^1,R^2, R^3 \}$, a representation of the Lie group $G^\ast_\k$ can be explicitly obtained as:
\be
G_\k^\ast  (\theta,p,\chi)=e^{\theta_3 D(R^3)} e^{\theta_2 D(R^2)} e^{\theta_1 D(R^1)} e^{p_1 D(X^1)} e^{p_2 D(X^2)} e^{p_3 D(X^3)} e^{\chi_1 D(L^1)} e^{\chi_2 D(L^2)} e^{\chi_3 D(L^3)} e ^{p_0 D(X^0)}.
\label{expads}
\ee
Moreover, a long but straightforward computation shows that if we multiply two $G_\k^\ast $ elements:
$$
G_\k^\ast  (\theta'',p'',\chi'')=G_\k^\ast  (\theta,p,\chi) \cdot G_\k^\ast  (\theta',p',\chi'),
$$
the group law 
$$
\theta''=f(\theta,\theta',p,p',\chi,\chi'),
\qquad
p''=g(\theta,\theta',p,p',\chi,\chi'),
\qquad
\chi''=h(\theta,\theta',p,p',\chi,\chi'),
$$
can be explicitly obtained and it can be exactly written as the coproduct~\eqref{rot}-\eqref{boosts} for $\k>0$, provided the identification
\be
\theta_a\equiv J_a, \qquad
p_0\equiv P_0, \qquad
p_a\equiv P_a, \qquad
\chi_a\equiv K_a,
\label{id31ptc}
\ee
is assumed and by following the convention~\eqref{alg}. Similarly to what observed for the $\omega=0$ case, the fact that we are able to recover the coproducts of the algebra in the bicrosscroduct basis is due to the specific choice of ordering of the exponential in \eqref{expads}. A different ordering choice would reflect into a group law compatible with the coproduct of the algebra in a different basis. We recall that the ordering~\eqref{expads} was just the one introduced in~\cite{BHMNplb2017} in order to obtain the $\kappa$-deformation of the (3+1) AdS algebra presented in Section 3, and this choice guarantees the self-consistency of all the new results here presented.

Now, the $\kappa$-AdS momentum space can be constructed by considering the left action of the group element $G_\k^\ast  (\theta,p,\chi)$ on an 8-dimensional ambient space. The points that can be reached from the origin ${\cal O}\equiv(0,0,0,0,0,0,0,1)$ under such action are those with coordinates
$(S_0,S_1,S_2,S_3,S_4,S_5,S_6,S_7)$ given by:
$$
G_\k^\ast\cdot
(0,0,0,0,0,0,0,1)^T=(S_0,S_1,S_2,S_3,S_4,S_5,S_6,S_7)^T\,.
\label{actionAdS}
$$
These can explicitly be written as:
\begin{align}
S_0&= \sinh(z p_0)+\frac12 e^{z p_0} z^2 \left( \bar{p}^2 - \omega \bar{\chi}^2 \right),\nonumber \\
S_1&= A \left(p_1 \; B_{21}^+ + \sqrt \omega \left( C + \chi_2 \; B_{21}^- \right) \right) ,\nonumber \\
S_2&= A \left( p_2 \; B_{12}^+ + \sqrt \omega \left( D - \chi_1 \; B_{12}^- \right) \right) ,\nonumber \\
S_3&= z e^{z p_0} \left( p_3 -z \sqrt \omega \left( \theta_1 \; p_1 + \theta_2 \; p_2 + \sqrt \omega \left( \theta_1 \; \chi_2 - \theta_2 \; \chi_1 \right) \right) \right), \nonumber \\
S_4&= A \left( -p_2 \; B_{21}^- + \sqrt \omega \left( D + \chi_1 \; B_{21}^+ \right) \right), \label{adsaction} \\
S_5&= A \left( -p_1 \; B_{12}^- + \sqrt \omega \left( C - \chi_2 \; B_{12}^+ \right) \right), \nonumber \\
S_6&= - z\,\sqrt \omega \, e^{z p_0} \left( \chi_3 -z \left( \theta_2 \; p_1 - \theta_1 \; p_2 + \sqrt \omega \left( \theta_1 \; \chi_1 + \theta_2 \; \chi_2 \right) \right) \right), \nonumber \\
S_7&= \cosh(z p_0)-\frac12 e^{z p_0} z^2 \left( \bar{p}^2 - \omega \bar{\chi}^2 \right),\nonumber
\end{align}
where we have defined: 
\begin{align}
\bar{p}^2&=p_1^2+p_2^2+p_3^2, \nonumber \\
\bar{\chi}^2&=\chi_1^2+\chi_2^2+\chi_3^2, \nonumber \\
A&=\frac12 z \,e^{z(p_0-\theta_3 \sqrt \omega)}, \nonumber \\
B_{ij}^\pm&= \omega \; z^2 \; (\theta_i^2-\theta_j^2)+e^{2 z \sqrt \omega \theta_3} \pm 1, \;\; i \in \{1,2\}, \label{auxss} \\
C&=2 z \left( \theta_2 \sqrt \omega \left( z \; \theta_1 \left( -p_2 + \sqrt \omega \; \chi_1 \right) - \chi_3 \right) + \theta_1 \; p_3 \right), \nonumber \\
D&=2 z \left( \theta_1 \sqrt \omega \left( z \; \theta_2 \left( -p_1 - \sqrt \omega \; \chi_2 \right) + \chi_3 \right) + \theta_2 \; p_3 \right).\nonumber
\end{align}
Note that, when evaluated at $(\theta_1, \theta_2, \theta_3 )=(0,0,0)$,  the last four functions  give $A \rightarrow \frac12 z e^{z p_0}$, $B_{ij}^+ \rightarrow 2$, $B_{ij}^- \rightarrow 0$, $C \rightarrow 0$ and $D \rightarrow 0$. 

We would like to  stress that the $\k\to 0$ limit of these expressions makes $S_4, S_5$ and $S_6$ vanish, and for the remaining ambient coordinates we get exactly the $\kappa$-Poincar\'e curved momentum space~\eqref{ambkp}. In other words, this means that for $\k=0$ the matrix~\eqref{expads} is a reducible representation of the dual $\kappa$-Poincar\'e group, which is consistent with the fact that the ambient space has been enlarged when the cosmological constant has been introduced.

From~\eqref{adsaction} we can deduce the geometrical properties of the $\kappa$-AdS momentum space. In fact, it is straightforward to check that the following relations hold:
\be
 -S_0^2 + S_1^2+ S_2^2 + S_3^2 - S_4^2 - S_5^2 -S_6^2 + S_7^2 = 1, 
 \qquad
 S_0 + S_7 =e^{z\,p_0} > 0 .
 \label{kadsms}
\ee
This means that, if we consider an  $\mathbb R^{4,4}$ ambient space, the $\kappa$-AdS momentum space is (half of) a  ${\rm  SO}(4,4)$ quadric. From the expressions~\eqref{adsaction} it is also straightforward to check that the subgroup of dual rotations,
\be
G_0=e^{\theta_3 D(R^3)} e^{\theta_2 D(R^2)} e^{\theta_1 D(R^1)}\,,
\label{dualr}
\ee
 leaves  the point ${\cal O}$ invariant. The action of the remaining 7-parameter subgroup (generated by the Lie subalgebra~\eqref{genkM}) is obtained by evaluating~\eqref{adsaction} at $(\theta_1, \theta_2, \theta_3 )=(0,0,0)$:
\be
\begin{aligned}
S_0&=\sinh(z p_0)+\frac12 \, e^{z p_0} z^2 \left( \bar{p}^2 - \omega \bar{\chi}^2 \right),
\\
S_1&=z\, e^{z p_0} \, p_1,
\\
S_2&=z\, e^{z p_0} \, p_2 ,
\\
S_3&=z\, e^{z p_0} \, p_3 ,
\\
S_4&=z\,  e^{z p_0} \,  \sqrt{\omega } \,\chi_1,
\\
S_5&=-z\,  e^{z p_0} \,  \sqrt{\omega } \,\chi_2 ,
\\
S_6&=-z \, e^{z p_0} \, \sqrt{\omega } \,\chi_3 ,
\\
S_7&=\cosh(z p_0)-\frac12\, e^{z p_0} z^2 \left( \bar{p}^2 - \omega \bar{\chi}^2 \right).
\end{aligned}
\label{AdSactionred}
\ee
These expressions encode the essential information concerning the non-vanishing cosmological constant generalization of~\eqref{ambkp}, since dual rotations leave the point ${\cal O}$ invariant. Therefore, we can think of the $\kappa$-AdS momentum space~\eqref{kadsms} as the 7-dimensional orbit in  $\mathbb R^{4,4}$ that can be parametrized through~\eqref{AdSactionred} in terms of the dual translation and boost coordinates, while the dual rotation coordinates $\theta$ do not play any role in the description of the curved momentum space.

We recall that the deformed Poisson brackets for the $\kappa$-AdS algebra would be the ones in~\cite{BHMNplb2017} for $\k>0$, and~\eqref{id31ptc} allows them to be interpreted as a Poisson-Lie structure on the dual Lie group $G_\k^\ast$ for which the multiplication on $G_\k^\ast$ ({\em i.e.} the coproduct~\eqref{rot}-\eqref{boosts}  for the $\kappa$-AdS algebra) is a Poisson map. If we now apply the identification~\eqref{id31ptc} onto the deformed Casimir function~\eqref{casa} and afterwards we project it onto the curved momentum space parametrized by the $p$ and $\chi$ coordinates by setting $\theta_i\to0$, 
we obtain\;
\be
{\cal C}_z =\frac 2{z^2}\left( \cosh (z\,p_0)-1 \right)-e^{z\,p_0} \left( \bar{p}^2 +\k \bar{\chi}^2 \right),
\label{casproj}
\ee
which could be considered as the deformed dispersion relation that corresponds to the (3+1) $\kappa$-AdS momentum space.\footnote{As we show in the following section, for $\omega<0$ this same expression describes the deformed dispersion relation corresponding to the (3+1) $\kappa$-dS momentum space.} The physical interpretation of such a dispersion relation requires to give a physical meaning to the hyperbolic momenta $\chi_{a}$. This can be done  thanks to the  identification \eqref{id31ptc}, which states that the local group coordinates have the same Poisson brackets as the generators of the $\kappa$-(A)dS algebra, as discussed in  detail in \cite{BGGHplb2017}. It is then possible to represent the local coordinates of the dual group  in terms of the usual phase space coordinates $\pi_{\mu}$ and $x^{\nu}$ with $\{\pi_{\mu}, x^{\nu}\}=\delta^\nu_\mu$, following a procedure similar to that found in \cite{Amelino-Camelia:2013uya, Barcaroli:2015eqe}. In particular, the $\chi_{a}$ are expected to be given by a combination of both the components $\pi_{\mu}$ and $x^{\nu}$ of the phase space. An explicit example of this is found in \cite{Barcaroli:2015eqe}, where the (1+1) $\kappa$-de Sitter algebra is represented on phase space.\footnote{Note that  the algebra appearing in \cite{Barcaroli:2015eqe} is written in a different basis than the one used here. While both algebras are of bicrossproduct type, for $z\rightarrow 0$ the algebra in \cite{Barcaroli:2015eqe} reduces to the de Sitter algebra in the comoving basis, while the one used here reduces to the  de Sitter algebra in kinematical basis. It is worth stressing that the comoving momenta are defined as a linear superposition of kinematical translations and boosts, which is again consistent the mixing of both type of kinematical transformations within~\eqref{casproj}.} In that case it is found that the dispersion relation depends on both momenta and spatial coordinates in a way that encodes a deformed gravitational redshift. We would expect something similar to show in this higher-dimensional model, however we are not yet at the stage where the explicit dependence of the dispersion relation on the phase space coordinates can be exposed. In fact, finding the appropriate phase space realization of the (3+1) $\kappa$-(A)dS algebra is a nontrivial task, that is work in progress and will be presented in a forthcoming publication~\cite{representation}, along with a thorough analysis of the physical implications of the findings presented here.


\section{The $\kappa$-dS curved momentum space}\label{dSmomspace}

As it could be expected, if we apply the construction presented in the previous Section to the case $\k<0$ we  obtain the same kind of geometric construction for the $\kappa$-dS momentum space, that should generalize the (2+1) results presented in~\cite{BGGHplb2017}. The only aspect we have to be careful about is the appearance of complex quantities when $\omega<0$, due to the presence of $\sqrt{\k}$ in some of the expressions (for instance, see~\eqref{ac}). This is not a major obstacle to the construction of the momentum space, since, as we are going to show,  all the complex contributions are linked to the dual of the rotation subgroup, which is again the isotropy subgroup of the origin of the momentum space. So they disappear when we consider the  projection to the submanifold parametrized by momenta and boost coordinates.

The matrix representation of the algebra~\eqref{ba} when $\k<0$ can be found to be
{\small
\be
D(X^0)=
z \left(
\begin{array}{cccccccc}
 0 & 0 & 0 & 0 & 0 & 0 & 0 & 1 \\
 0 & 0 & 0 & 0 & 0 & 0 & 0 & 0 \\
 0 & 0 & 0 & 0 & 0 & 0 & 0 & 0 \\
 0 & 0 & 0 & 0 & 0 & 0 & 0 & 0 \\
 0 & 0 & 0 & 0 & 0 & 0 & 0 & 0 \\
 0 & 0 & 0 & 0 & 0 & 0 & 0 & 0 \\
 0 & 0 & 0 & 0 & 0 & 0 & 0 & 0 \\
 1 & 0 & 0 & 0 & 0 & 0 & 0 & 0 \\
\end{array}
\right),
\qquad
D(X^1)=
z \left(
\begin{array}{cccccccc}
 0 & 1 & 0 & 0 & 0 & 0 & 0 & 0 \\
 1 & 0 & 0 & 0 & 0 & 0 & 0 & 1 \\
 0 & 0 & 0 & 0 & 0 & 0 & 0 & 0 \\
 0 & 0 & 0 & 0 & 0 & 0 & 0 & 0 \\
 0 & 0 & 0 & 0 & 0 & 0 & 0 & 0 \\
 0 & 0 & 0 & 0 & 0 & 0 & 0 & 0 \\
 0 & 0 & 0 & 0 & 0 & 0 & 0 & 0 \\
 0 & -1 & 0 & 0 & 0 & 0 & 0 & 0 \\
\end{array}
\right),
\nonumber
\ee
\be
D(X^2)=
z \left(
\begin{array}{cccccccc}
 0 & 0 & 1 & 0 & 0 & 0 & 0 & 0 \\
 0 & 0 & 0 & 0 & 0 & 0 & 0 & 0 \\
 1 & 0 & 0 & 0 & 0 & 0 & 0 & 1 \\
 0 & 0 & 0 & 0 & 0 & 0 & 0 & 0 \\
 0 & 0 & 0 & 0 & 0 & 0 & 0 & 0 \\
 0 & 0 & 0 & 0 & 0 & 0 & 0 & 0 \\
 0 & 0 & 0 & 0 & 0 & 0 & 0 & 0 \\
 0 & 0 & -1 & 0 & 0 & 0 & 0 & 0 \\
\end{array}
\right),
\qquad
D(X^3)=
z \left(
\begin{array}{cccccccc}
 0 & 0 & 0 & 1 & 0 & 0 & 0 & 0 \\
 0 & 0 & 0 & 0 & 0 & 0 & 0 & 0 \\
 0 & 0 & 0 & 0 & 0 & 0 & 0 & 0 \\
 1 & 0 & 0 & 0 & 0 & 0 & 0 & 1 \\
 0 & 0 & 0 & 0 & 0 & 0 & 0 & 0 \\
 0 & 0 & 0 & 0 & 0 & 0 & 0 & 0 \\
 0 & 0 & 0 & 0 & 0 & 0 & 0 & 0 \\
 0 & 0 & 0 & -1 & 0 & 0 & 0 & 0 \\
\end{array}
\right),
\nonumber
\ee
\be
D(L^1)=
z\,\sqrt{-\omega }  \left(
\begin{array}{cccccccc}
 0 & 0 & 0 & 0 & 1 & 0 & 0 & 0 \\
 0 & 0 & 0 & 0 & 0 & 0 & 0 & 0 \\
 0 & 0 & 0 & 0 & 0 & 0 & 0 & 0 \\
 0 & 0 & 0 & 0 & 0 & 0 & 0 & 0 \\
 1 & 0 & 0 & 0 & 0 & 0 & 0 & 1 \\
 0 & 0 & 0 & 0 & 0 & 0 & 0 & 0 \\
 0 & 0 & 0 & 0 & 0 & 0 & 0 & 0 \\
 0 & 0 & 0 & 0 & -1 & 0 & 0 & 0 \\
\end{array}
\right),
\qquad
D(L^2)=
z\,\sqrt{-\omega }  \left(
\begin{array}{cccccccc}
 0 & 0 & 0 & 0 & 0 & 1 & 0 & 0 \\
 0 & 0 & 0 & 0 & 0 & 0 & 0 & 0 \\
 0 & 0 & 0 & 0 & 0 & 0 & 0 & 0 \\
 0 & 0 & 0 & 0 & 0 & 0 & 0 & 0 \\
 0 & 0 & 0 & 0 & 0 & 0 & 0 & 0 \\
 1 & 0 & 0 & 0 & 0 & 0 & 0 & 1 \\
 0 & 0 & 0 & 0 & 0 & 0 & 0 & 0 \\
 0 & 0 & 0 & 0 & 0 & -1 & 0 & 0 \\
\end{array}
\right),
\nonumber
\ee
\be
D(L^3)=
z\,\sqrt{-\omega }  \left(
\begin{array}{cccccccc}
 0 & 0 & 0 & 0 & 0 & 0 & 1 & 0 \\
 0 & 0 & 0 & 0 & 0 & 0 & 0 & 0 \\
 0 & 0 & 0 & 0 & 0 & 0 & 0 & 0 \\
 0 & 0 & 0 & 0 & 0 & 0 & 0 & 0 \\
 0 & 0 & 0 & 0 & 0 & 0 & 0 & 0 \\
 0 & 0 & 0 & 0 & 0 & 0 & 0 & 0 \\
 1 & 0 & 0 & 0 & 0 & 0 & 0 & 1 \\
 0 & 0 & 0 & 0 & 0 & 0 & -1 & 0 \\
\end{array}
\right),
\qquad
D(R^1)=
z\,\sqrt{-\omega } \left(
\begin{array}{cccccccc}
 0 & 0 & 0 & 0 & 0 & 0 & 0 & 0 \\
 0 & 0 & 0 & i & 0 & 0 & 0 & 0 \\
 0 & 0 & 0 & 0 & 0 & 0 & -1 & 0 \\
 0 & -i & 0 & 0 & 0 & 1 & 0 & 0 \\
 0 & 0 & 0 & 0 & 0 & 0 & i & 0 \\
 0 & 0 & 0 & -1 & 0 & 0 & 0 & 0 \\
 0 & 0 & 1 & 0 & -i & 0 & 0 & 0 \\
 0 & 0 & 0 & 0 & 0 & 0 & 0 & 0 \\
\end{array}
\right),
\nonumber
\ee
\be
D(R^2)=
z\,\sqrt{-\omega }  \left(
\begin{array}{cccccccc}
 0 & 0 & 0 & 0 & 0 & 0 & 0 & 0 \\
 0 & 0 & 0 & 0 & 0 & 0 & 1 & 0 \\
 0 & 0 & 0 & i & 0 & 0 & 0 & 0 \\
 0 & 0 & -i & 0 & -1 & 0 & 0 & 0 \\
 0 & 0 & 0 & 1 & 0 & 0 & 0 & 0 \\
 0 & 0 & 0 & 0 & 0 & 0 & i & 0 \\
 0 & -1 & 0 & 0 & 0 & -i & 0 & 0 \\
 0 & 0 & 0 & 0 & 0 & 0 & 0 & 0 \\
\end{array}
\right),
\qquad
D(R^3)=
z\,\sqrt{-\omega }  \left(
\begin{array}{cccccccc}
 0 & 0 & 0 & 0 & 0 & 0 & 0 & 0 \\
 0 & 0 & 0 & 0 & 0 & -1 & 0 & 0 \\
 0 & 0 & 0 & 0 & 1 & 0 & 0 & 0 \\
 0 & 0 & 0 & 0 & 0 & 0 & 0 & 0 \\
 0 & 0 & -1 & 0 & 0 & 0 & 0 & 0 \\
 0 & 1 & 0 & 0 & 0 & 0 & 0 & 0 \\
 0 & 0 & 0 & 0 & 0 & 0 & 0 & 0 \\
 0 & 0 & 0 & 0 & 0 & 0 & 0 & 0 \\
\end{array}
\right).
\nonumber
\ee
}

Again, the left linear action of~\eqref{expads} onto the  point ${\cal O}=(0,0,0,0,0,0,0,1)$ gives rise to an orbit whose  points have ambient coordinates in $R^{1,7}$ given by:
\begin{align}
S_0&= \sinh(z p_0)+\frac12 e^{z p_0} z^2 \left( \bar{p}^2 - \omega \bar{\chi}^2 \right),\nonumber \\
S_1&= A \left(p_1 \; B_{21}^+ + i \sqrt{-\omega} \left( C + \chi_2 \; B_{21}^- \right) \right), \nonumber \\
S_2&= A \left( p_2 \; B_{12}^+ + i \sqrt{-\omega} \left( D - \chi_1 \; B_{12}^- \right) \right) ,\nonumber \\
S_3&= z e^{z p_0} \left( p_3 -i \; z \sqrt {-\omega} \left( \theta_1 \; p_1 + \theta_2 \; p_2 + i \sqrt{-\omega} \left( \theta_1 \; \chi_2 - \theta_2 \; \chi_1 \right) \right) \right), \nonumber \\
S_4&= A \left( i \; p_2 \; B_{21}^- + \sqrt{-\omega} \left( D + \chi_1 \; B_{21}^+ \right) \right), \label{actiondS} \\
S_5&= A \left( -i \; p_1 \; B_{12}^- - \sqrt{-\omega} \left( C - \chi_2 \; B_{12}^+ \right) \right), \nonumber \\
S_6&= \sqrt{-\omega} \; z \; e^{z p_0} \left( \chi_3 -z \left( \theta_2 \; p_1 - \theta_1 \; p_2 + i \; \sqrt{-\omega} \left( \theta_1 \; \chi_1 + \theta_2 \; \chi_2 \right) \right) \right), \nonumber \\
S_7&= \cosh(z p_0)-\frac12 e^{z p_0} z^2 \left( \bar{p}^2 - \omega \bar{\chi}^2 \right), \nonumber
\end{align}
where $A, B_{ij}^\pm, C, D$ are the same functions appearing in~\eqref{auxss}. It is straightforward to check that such coordinates obey the constraints:
$$
-S_0^2 + S_1^2+ S_2^2 + S_3^2 + S_4^2 + S_5^2 + S_6^2 + S_7^2 = 1, 
\qquad
S_0 + S_7 = e^{z\,p_0} > 0,
$$
so that  we obtain (half of) the (6+1) dS space as the curved momentum space for the $\kappa$-dS quantum algebra.

Again, the isotropy subgroup for ${\cal O}$ is generated by the subgroup of dual rotations~\eqref{dualr}, and each point of the curved momentum space can be charaterized by the seven momenta and rapidities by evaluating~\eqref{actiondS}  at $(\theta_1, \theta_2, \theta_3 )=(0,0,0)$:
\be
\begin{aligned}
S_0&=\sinh(z p_0) + \frac{1}{2}\, e^{z p_0} z^2 \left(\bar{p}^2 -\omega  \bar{\chi}^2 \right),
\nonumber\\
S_1&=z\,e^{z p_0 } \,p_1 ,
\nonumber\\
S_2&=z \, e^{z p_0 } \,p_2,
\nonumber\\
S_3&=z\, e^{z p_0 } \,p_3 ,
\nonumber\\
S_4&=z  \, e^{z p_0} \,\sqrt{-\omega }\,\chi_1,\\
S_5&=z\,e^{z p_0} \,\sqrt{-\omega }\,\chi_2 ,\\
S_6&=z\, e^{z p_0} \,\sqrt{-\omega }\,\chi_3 ,\\
S_7&=\cosh(z p_0) - \frac{1}{2} e^{z p_0} \, z^2 \left(\bar{p}^2 -\omega  \bar{\chi}^2 \right)\,.
\end{aligned}
\label{sproyds}
\ee

Note that these ambient space coordinates~\eqref{sproyds} are all real, since all the complex contributions in~\eqref{actiondS} are linked to the action of the dual rotation subgroup.

Finally, the projection of the deformed Casimir onto the curved momentum space leads  to~\eqref{casproj}, 
which can again be  interpreted as the dispersion relation for the $\k<0$ case. Also, the $\k\to 0$ Poincar\'e limit of all of these expressions is straightforward, and leads to the results presented in Section 2.

\section{Discussion and concluding remarks}\label{conclusions}

In  \cite{BGGHplb2017} we constructed the generalized momentum space associated to quantum deformed spacetime symmetries in presence of a cosmological constant.
We focused on the $\kappa$-dS algebra, which can be seen as a deformation of the standard Poincar\'e algebra governed by two deformation parameters: the cosmological constant $\Lambda=-\omega$ is a classical deformation parameter, controlling the effects of spacetime curvature, while $z=1/\kappa$ is the quantum deformation parameter, controlling the Planck-scale effects (which in turn induce curvature in momentum space).
The interplay between these two deformations is nontrivial and intertwines the coordinates of the whole phase space. So, while it was known for a while already that in the $\omega\rightarrow 0$ limit (where the symmetries are described by the $\kappa$-Poincar\'e algebra) the geometrical properties of the momentum space are those of half of a dS manifold, it was generally thought that it would not be possible to make an analogous analysis once $\omega\neq 0$.

Our recent result was  an important breakthrough in this respect, since it demonstrated that the joint effects of spacetime curvature and of the quantum deformation can be taken into account if one constructs  the momentum space by considering not just the momenta linked to spacetime translations, but also  the hyperbolic momenta associated to boost transformations. Still,  \cite{BGGHplb2017} focussed only on low-dimensional cases, namely the $\kappa$-dS algebras in (1+1) and (2+1) dimensions. Of course the physically relevant model is that with (3+1) dimensions, which, as explained in the introduction, allows to connect to the phenomenology of particles propagating over cosmological distances. 
Describing the generalized momentum space of both quantum-deformed dS and AdS algebras with a cosmological constant in (3+1) dimensions was the goal of the work presented here. 

Going from the (2+1)-dimensional case to the one with (3+1) dimensions entails dealing with a deformed rotation sector, which is still classical in lower dimensional models. Specifically, the coalgebra of the rotations is modified in (3+1) dimensions, in such a way that one of the rotation generators  takes a special role compared to the others (see the Appendix). This might raise worries that the model breaks spatial isotropy. However, just as the deformed boost transformations do not break relativistic invariance, but simply deform the laws of transformation between inertial frames, the deformed rotations  could imply that  the concept of isotropy has to be adapted to fit within the new transformation rules. What the observational consequences of this deformed isotropy could be is still a matter of investigation.

Despite these novel features, the analysis of the generalized momentum space of the $\kappa$-dS algebra in (3+1) dimensions led to a higher-dimensional version of the results found in \cite{BGGHplb2017}: the momentum space is half of a (6+1)-dimensional dS manifold and the rotations are the group of isotropy of its origin. The lower-dimensional results are recovered  via canonical projection. 

We expanded our construction to the $\kappa$-AdS algebra, which can be defined starting from the $\kappa$-dS algebra and changing sign to the cosmological constant parameter. While the difference between the two models is minimal at the level of the algebra and coalgebra, we found that the momentum space is characterized by a qualitatively different geometry. This is because the change of sign of the cosmological constant produces the appearance of complex quantities due to the presence of $\sqrt{\omega}$ factors.  This analysis provides the first example where quantum effects do not produce a momentum space with dS geometry, but something different --we found that the $\kappa$-AdS algebra has a momentum manifold with ${\rm SO}(4,4)$ invariance.

Finally, we would like to comment that the non-relativistic limit of the results here presented is indeed worth to be studied, since it would give rise to the momentum spaces associated to quantum deformations of the (3+1) Newton-Hooke symmetries when $\k\neq 0$, and to quantum Galilei symmetries in the case $\k=0$. In this respect, we recall that the Galilean limit of (2+1) quantum gravity based on quantum deformations of the Galilei and Newton-Hooke algebras was formerly presented in~\cite{PSchroers1} and~\cite{PSchroers2}, and further studied in~\cite{pedro}. Work on this line is in progress.


\section*{Acknowledgments}

A.B., I.G-S. and F.J.H. have been partially supported by Ministerio de Econom\'{i}a y Competitividad (MINECO, Spain) under grants MTM2013-43820-P and   MTM2016-79639-P (AEI/FEDER, UE), by Junta de Castilla y Le\'on (Spain) under grants BU278U14 and VA057U16 and by the Action MP1405 QSPACE from the European Cooperation in Science and Technology (COST). I.G-S. acknowledges a PhD grant from the Junta de Castilla y Le\'on (Spain) and European Social Fund. G.G. acknowledges a Grant for Visiting Researchers at the Campus of International Excellence `Triangular-E3' (MECD, Spain).


\section*{Appendix}
\setcounter{equation}{0}
\renewcommand{\theequation}{A.\arabic{equation}}

Let us recall that since $\delta$ in Eq.~\eqref{ac} is a coboundary Lie bialgebra, there exists a classical $r$-matrix such that $\delta(X)=[ X \otimes 1+1\otimes X ,  r]$.  For coboundary deformations classical $r$-matrices provide the simplest way to condense the information of a given quantum algebra. In this case (see~\cite{BHMNplb2017,LBC}) such classical $r$-matrix reads
\be
r_\k=z \left( K_1 \wedge P_1 + K_2 \wedge P_2+ K_3 \wedge P_3 + \sqrt{\k}J_1 \wedge J_2 \right),
\label{ab}
\ee
and its $\kappa$-Poincar\'e limit $\k\to 0$ is the well-known classical $r$-matrix
\be
r=z \left( K_1 \wedge P_1 + K_2 \wedge P_2+ K_3 \wedge P_3 \right).
\label{abkp}
\ee
Note that rotations appear in the $r$-matrix only in (3+1) dimensions. In fact, the (2+1) $\kappa$-\adsw deformations are generated by the classical $r$-matrix given by (see~\cite{tallin,BHMNsigma})
\be
r=z \left( K_1 \wedge P_1 + K_2 \wedge P_2 \right),
\label{ab21}
\ee
which does not depend on $\k$ and is just the projection of~\eqref{ab} to the (2+1)-dimensional case, obtained by supressing the $J_1,J_2,P_3$ and $K_3$ generators. The $J_1 \wedge J_2$ term, which does not appear in  \eqref{ab21},  is  the one responsible for the non-vanishing $\delta(J)$ in the (3+1) case and, therefore, is the term that induces the deformation~\eqref{rot} of the rotation subalgebra.

At this point it is natural to wonder whether there could exist another quantum \adsw algebra in (3+1) dimensions that generalizes the $\kappa$-Poincar\'e algebra and has a non-deformed rotation subalgebra, $\delta(J_1)=\delta(J_2)=\delta(J_3)=0$. This  question can be addressed by recalling that in~\cite{prague,tallin} it was proven that all \adsw deformations in (3+1) dimensions with primitive coproducts for $P_0$ and $J_3$ ({\em i.e.} with $\delta(P_0)=\delta(J_3)=0$) are generated by one of the  two (disjoint) families of  two-parametric classical $r$-matrices:
     \bea
     &&\!\!\!\!\!\!\!\!\!\!\!\! r_{\z_1,\z_3}=\z_1\left(K_1\wedge
P_1+K_2\wedge P_2+K_3\wedge P_3 \pm  \sqrt{\k}\,  J_1\wedge
J_2\right) +\z_3 P_0\wedge J_3    ,\label{r1}\\ &&\!\!\!\!\!\!\!\!\!\!\!\!
r_{\z_2,\z_3}= \z_2\left(P_1\wedge P_2+\k K_1\wedge K_2-\k\,
J_1\wedge J_2\pm\sqrt{\k}  P_3\wedge K_3\right)  +\z_3 P_0\wedge J_3   .
\label{r2}
\eea
Therefore, if we impose that the $\kappa$-Poincar\'e algebra has to be obtained in the limit $\omega\to 0$ (which means that  the classical $r$-matrix that generates the deformation should still have~\eqref{abkp} as its $\k\to 0$ limit), this implies that the only viable solution  is \eqref{r1}, that is,~\eqref{ab} plus  an additional twist generated by $\z_3 P_0\wedge J_3$.  We point out  that both classical $r$-matrices (\ref{r1}) and (\ref{r2}) are invariant under the automorphism (\ref{auto})  provided that the deformation parameters $z_1$, $z_2$ and $z_3$ remain unchanged.

It can also be  proven~\cite{tallin} that a  linear change of basis $X\rightarrow \hat X$ in the $\kappa$-\adsw algebra transforms the $r$-matrix~\eqref{r1} into 
\bea
&&\!\!\!\!\!\!\!\!\!\!\!\!\!\!\!\! r_{\z_1,\hat\z_3}=\z_1\left(\hat
K_1\wedge \hat P_1+\hat K_2\wedge\hat P_2+\hat K_3\wedge \hat P_3 \pm 
\frac{\sqrt{\k}}{\sqrt{3}}\,\left( \hat J_1\wedge \hat J_2+ \hat
J_2\wedge \hat J_3+\hat J_3\wedge\hat  J_1\right)\right)\cr  &&
\qquad\qquad +\hat\z_3\hat P_0\wedge \left(\hat J_1+\hat J_2+\hat J_3 
\right)  ,
\nonumber
\eea
where $
\hat z_3=  z_3/\sqrt{3}
$, in such a way that the three rotation generators seem to play the same role, although now it can be checked that the primitive generator for the rotation subalgebra turns out to be $\hat J_1+\hat J_2+\hat J_3$. As a consequence, we can state as a general result that the generalization of the $\kappa$-Poincar\'e algebra to the non-vanishing cosmological constant case implies a non-trivial Planck scale deformation of the rotation sector, unless we are willing to deform the time-translation sector.


 \end{document}